# Discovery of a Single-Band Mott Insulator in a van der Waals Flat-Band Compound

Shunye Gao,[1,2,3,#] Shuai Zhang,[1,2,#] Cuixiang Wang,[1,2,#] Shaohua Yan,[4,#] Xin Han,[1,2], Xuecong Ji,[1,2] Wei Tao,[1] Jingtong Liu,[1] Tiantian Wang,[1] Shuaikang Yuan,[1] Gexing Qu,[1,2] Ziyan Chen,[1,2] Yongzhao Zhang,[1,2] Jierui Huang,[1,2] Mojun Pan,[1,2] Shiyu Peng,[1,2] Yong Hu,[3] Hang Li,[3] Yaobo Huang,[5] Hui Zhou,[1,2] Sheng Meng,[1,2,6] Liu Yang,[7,8] Zhiwei Wang,[7,8,9] Yugui Yao,[7,8] Zhiguo Chen,[1,6] Ming Shi,[3,10] Hong Ding,[1,2,6] Huaixin Yang,[1,2,6] Kun Jiang,[1,2,6] Yunliang Li,[1,2,6] Hechang Lei,[4,*] Youguo Shi,[1,2,6,*] Hongming Weng,[1,2,6,*], and Tian Qian,[1,6,*]

[1] *Beijing National Laboratory for Condensed Matter Physics and Institute of Physics, Chinese Academy of Sciences, Beijing, 100190, China*

[2] *University of Chinese Academy of Sciences, Beijing, 100049, China*

[3] *Photon Science Division, Paul Scherrer Institut, CH-5232 Villigen, Switzerland*

[4] *Department of Physics and Beijing Key Laboratory of Opto-electronic Functional Materials and Micro-nano Devices, Renmin University of China, Beijing, 100872, China*

[5] *Shanghai Synchrotron Radiation Facility, Shanghai Advanced Research Institute, Chinese Academy of Sciences, Shanghai, 201210, China*

[6] *Songshan Lake Materials Laboratory, Dongguan, Guangdong, 523808, China*

[7] *Centre of Quantum Physics, Key Laboratory of Advanced Optoelectronic Quantum Architecture and Measurement, Ministry of Education, School of Physics, Beijing Institute of Technology, Beijing, 100081, China*

[8] *Micronano Center, Beijing Key Lab of Nanophotonics and Ultrafine Optoelectronic Systems, Beijing Institute of Technology, Beijing, 100081, China*

[9] *Material Science Center, Yangtze Delta Region Academy of Beijing Institute of Technology, Jiaxing, 314011, China*

[10] *Center for Correlated Matter and Department of Physics, Zhejiang University, Hangzhou, 310058, China*

[#] These authors contributed equally to this work.

[*] Corresponding authors: tqian@iphy.ac.cn, hmweng@iphy.ac.cn, ygshi@iphy.ac.cn, hlei@ruc.edu.cn

**Abstract**

The Mott insulator provides an excellent foundation for exploring a wide range of strongly correlated physical phenomena, such as high-temperature superconductivity, quantum spin liquid, and colossal magnetoresistance. A Mott insulator with the simplest degree of freedom is an ideal and highly desirable system for studying the fundamental physics of Mottness. In this study, we have unambiguously identified such an anticipated Mott insulator in a van der Waals layered compound $Nb_3Cl_8$. In the high-temperature phase, where interlayer coupling is negligible, density functional theory calculations for the monolayer of $Nb_3Cl_8$ suggest a half-filled flat band at the Fermi level, whereas angle-resolved photoemission spectroscopy experiments observe a large gap. This observation is perfectly reproduced by dynamical mean-field theory calculations considering strong electron correlations, indicating a correlation-driven Mott insulator state. Since this half-filled band derived from a single $2a_1$ orbital is isolated from all other bands, the monolayer of $Nb_3Cl_8$ is an ideal realization of the celebrated single-band Hubbard model. Upon decreasing the temperature, the bulk system undergoes a phase transition, where structural changes significantly enhance the interlayer coupling. This results in a bonding-antibonding splitting in the Hubbard bands, while the Mott gap remains dominant. Our discovery provides a simple and seminal model system for investigating Mott physics and other emerging correlated states.

# I. INTRODUCTION

In solid-state physics, the classification of materials based on their transport properties is one of the oldest and least understood problems. According to modern band theory, metals possess partially filled bands, while insulators have all bands fully filled. However, this explanation is incomplete. The physics regarding partially filled bands faces significant challenges due to electron-electron correlations. One typical example is the formation of Mott insulators [1,2]. Mott physics is governed by two critical energy scales: the bandwidth ($W$) and the correlation strength ($U$). As illustrated in Fig. 1(a), for a half-filled system, if $U \ll W$, the system falls into the metallic band regime. Conversely, if $U \gg W$, the system transforms into a Mott insulator, where strong correlations spit the metallic band into two: The lower band is filled while the upper one is empty. In Mott insulators, localized electrons are expected to carry local magnetic moments, often resulting in an antiferromagnetic ground state, as observed in the parent compounds of cuprate superconductors [3]. In addition, Mott insulators exhibit various magnetic ground states, such as spin singlet state [4,5], quantum spin liquid state [6], and valence bond state [7]. Beyond this unconventional insulating state, numerous intriguing physical phenomena are triggered from Mott insulators, such as high-temperature superconductivity [3], colossal magnetoresistance [8,9], non-Fermi liquid [10], and so on. Many strongly correlated systems have been investigated within the framework of Mott physics, also known as Mottness, including the recently discovered insulator behavior and superconductivity in magic-angle twisted bilayer graphene [11,12].

The theoretical understanding for the Mott physics has been achieved using simplified models, in particular the single-band Hubbard model, which considers electrons on one band in a single orbital [1]. However, low-energy excitations in real materials are usually complicated by various factors, such as multiple bands, orbital degeneracy, spin-orbit coupling, Peierls instability, and so on [13], making experimental verification and theoretical analysis challenging and controversial. Well-known examples are the parent compounds of cuprate superconductors, which feature a half-filled band at the Fermi level ($E_F$), arising from a strong hybridization of the Cu $3d_{x^2-y^2}$ orbital with the O $2p_x$ and $2p_y$ orbitals. Consequently, the low-energy excitations are described by the Zhang-Rice singlets emerging from the strong *p-d* hybridization [14]. In the 5*d* system Sr$_2$IrO$_4$, the partially filled wide $t_{2g}$ states would lead to metallic behavior. Strong spin-orbit coupling is required to split the $t_{2g}$ states, resulting in a separated half-filled band with a relatively narrow width [15]. The metal-insulator transition in VO$_2$ is accompanied by a dimerization of the V atoms [4]. The origin of the insulating gap has been a subject of debate, whether it arises from bonding-antibonding splitting [16,17] or electron correlations [5,18-20]. In the charge-density-wave phase of 1*T*-TaS$_2$, band folding and hybridization result in a reconstructed flat band at $E_F$ in the superlattice [21,22]. However, the interplay between interlayer coupling and electron correlations complicates the insulating behavior in 1*T*-TaS$_2$ [23,24].

In this work, we have discovered a textbook example of the single-band Mott insulator in a van der Waals (vdW) compound Nb$_3$Cl$_8$. The electronic states in the high-temperature phase can be adequately described based on the monolayer of Nb$_3$Cl$_8$, as the interlayer coupling is negligible. Density functional theory (DFT) calculations show that the monolayer of Nb$_3$Cl$_8$ features a single half-filled band derived from the $2a_1$ orbital at $E_F$. This band exhibits a nearly flat dispersion with $W \sim 0.3$ eV, which is significantly smaller than the typical $U$ values on the order of eV for Nb 4*d* orbitals [25-28], indicating its strong correlation nature. Strong correlations drive the half-filled system to transform into a Mott insulator, resulting in a splitting of the half-filled band into lower and upper Hubbard bands (LHB and UHB). Since this half-filled $2a_1$ band is well-separated from other bands, the monolayer of Nb$_3$Cl$_8$ can be ideally described by the single-band Hubbard model. Moreover, the Curie-Weiss paramagnetism observed in the high-temperature phase indicates the existence of spin-1/2 local moments, which is in excellent agreement with the prediction of the single-band Hubbard model.

As the temperature decreases, the bulk system undergoes a phase transition. In the low-temperature phase, the structural changes in the stacking sequence greatly enhance the interlayer coupling, leading to significant impacts on the correlated electronic states. A bonding/antibonding splitting occurs in the Hubbard bands, while the spins pair into interlayer spin singlets, resulting in nonmagnetic behavior. Despite these remarkable changes, the system remains in the Mott insulator regime in the low-temperature phase, as the insulating gap is still dominated by strong electron correlations.

## II. INSULATING STATE ON A HALF-FILLED BAND

In the high-temperature phase ($\alpha$ phase) above $T^* \sim 100$ K, each unit cell consists of two monolayers of $Nb_3Cl_8$ stacked along the $c$-axis via weak vdW forces [Fig. 1(b)] [29,30]. In each monolayer, the Nb ions form a breathing kagome lattice, where three Nb ions are in close proximity, constituting one $Nb_3$ trimer [Fig. 1(c)] [29-31]. Each $Nb_3$ trimer is surrounded by 13 Cl ions, forming a $Nb_3Cl_{13}$ cluster composed of three edge-sharing $NbCl_6$ octahedra [inset of Fig. 1(d)]. The intra-cluster Nb–Nb distance $d_1$ = 2.834 Å is substantially shorter than the inter-cluster Nb–Nb distance $d_2$ = 3.998 Å [32], indicating the formation of Nb–Nb metal bonds within each cluster. The octahedral crystal field splits the Nb $4d$ orbitals into $e_g$ and $t_{2g}$ orbitals, which further divide into molecular orbitals due to the metal bonds, as illustrated in Fig. 1(d). The $3p$ shell of the Cl ions is filled due to their strong electronegativity. Consequently, the $Nb_3$ trimers in $Nb_3Cl_8$ have a valence state of +8, with seven valence electrons occupying the molecular orbitals. Six valence electrons occupy the lowest three molecular orbitals, while the seventh electron resides in the $2a_1$ orbital [29,30,33,34].

The occupied molecular orbitals correspond to four bands between -3 and 0.5 eV in the DFT calculations shown in Fig. 1(e). Because of symmetry breaking and large lattice distortion in the breathing kagome lattice, the bands deviate significantly from the characteristic band structure of the conventional kagome lattice. The bandwidths are determined by the hybridization between the Nb $4d$ orbitals and the $3p$ orbitals of the ligand Cl ions that bridge adjacent $Nb_3$ trimers. The nearly flat band dispersions signify an extremely weak *p-d* hybridization, in contrast to the strong *p-d* hybridization in cuprates [14]. According to band theory, each filled band can accommodate two electrons. As a result, the half-filled $2a_1$ band, which is occupied by one electron, is pinned at $E_F$ in the DFT calculations.

The band structure obtained within the single-particle approximation suggests that $\alpha$-Nb$_3$Cl$_8$ exhibits a metallic state with a half-filled band at $E_F$ [Fig. 2(a)]. In contrast, the angle-resolved photoemission spectroscopy (ARPES) results reveal a clear gap, as shown in Fig. 2(b). Despite performing the ARPES experiments at room temperature, the spectra are affected by charging effects. Specifically, the kinetic energy of photoelectrons decreases with increasing photon flux. The charging effects suggest that the samples exhibit highly insulating behavior, which is consistent with previous transport measurements [35]. By measuring the kinetic energy with different beamline slit sizes, which are proportional to photon flux, we determine that the valence band top at the Γ point is located 0.7 eV below $E_F$ [see Fig. S1 in Supplemental Material [36]].

Except the half-filled band, the bands resolved by ARPES measurements show excellent consistency with the DFT calculations over a wide energy range. The calculations in Fig. 2(a) reveal that the valence band structure consists of two well-separated sets of bands. One set originates from the Cl $3p$ orbitals and lies well below $E_F$ due to the strong electronegativity, while the other set, with lower binding energies, originates from the Nb $4d$ orbitals. The photon energy ($h\nu$)-dependent data in Fig. 2(c) show significant suppression of the spectral intensities of the bands within 3 eV below $E_F$ around $h\nu = 32$ eV, which corresponds to the binding energy of the Nb $4p$ orbitals (see Fig. S5(d) in Supplemental Material [36]). This behavior is attributed to the Nb $4p$-$4d$ antiresonance, similar to the Fe $3p$-$3d$ antiresonance observed in iron-based superconductors [37,38], confirming that these bands originate from the Nb $4d$ orbitals.

Although the ARPES results reveal a clear gap at $E_F$, they do not provide information on the gap size because the bands above $E_F$ cannot be detected in the ARPES experiments. To determine the gap size at $E_F$, we performed photoluminescence (PL) experiments. We note that PL spectra may also exhibit signals related to excitonic or impurity states. The PL spectrum excited by 1.1-eV photons exhibits two peaks at 0.7 and 1.0 eV, respectively [Fig. 2(d)]. However, the 0.7 eV peak vanishes when directly excited by 0.75-eV photons, as shown in Fig. 2(e). The results suggest that the 0.7 eV peak does not originate from intrinsic bands or impurity states. Instead, it is likely associated with an excitonic state formed by the Coulomb interactions between photoexcited electrons and holes during the optical excitation process. The 1.0 eV peak is attributed to the recombination of photoexcited electrons at the conduction band bottom and holes at the valence band top, indicating a gap size of 1.0 eV in $\alpha$-Nb$_3$Cl$_8$ [see Fig. S2(b) in Supplemental Material [36]].

## III. MOTT INSULATOR DESCRIBED BY SINGLE-BAND HUBBARD MODEL

The observed insulating state with a large gap at $E_\text{F}$ in $\alpha$-Nb$_3$Cl$_8$ contradicts the metallic state predicted by the DFT calculations within the single-particle approximation. In a previous report, the ARPES results also showed an insulating state, which was interpreted as a band insulator by assuming that $E_\text{F}$ lies within the band gap above the calculated half-filled band [39]. Under this assumption, this band should be fully occupied by two electrons, resulting in a total of eight valence electrons filling four Nb 4$d$ bands. However, this interpretation contradicts the fact that each [Nb$_3$]$^{8+}$ trimer possesses seven valence electrons.

To elucidate the origin of the insulating state, we extract the experimental band dispersions and compare them with the band structure calculated using the Heyd-Scuseria-Ernzerhof (HSE06) hybrid functional in Fig. 3(a). Four bands are identified within 3 eV below $E_\text{F}$ in the ARPES data [see Figs. S5(a) and S5(b) in Supplemental Material [36]]. The lower three bands exhibit excellent consistency with the DFT calculations, signifying an accurate description within the single-particle approximation. However, a notable discrepancy arises in the topmost band between experiment and calculation. The calculations suggest a half-filled 2$a_1$ band at $E_\text{F}$, whereas the ARPES data show it well below $E_\text{F}$. In addition, the calculated 2$a_1$ band needs to be renormalized by a factor of 1.6 to match the experimental data [see Fig. S5(c) in Supplemental Material [36]], highlighting the importance of strong correlation effects on this band beyond the single-particle approximation. The half-filled band has a nearly flat dispersion with $W \sim 0.3$ eV, while typical $U$ values on Nb 4$d$ orbitals are on the order of eV [25-28]. Given that $U$ is much larger than $W$ on this half-filled band, it is thus reasonable to expect a correlation-driven Mott transition to occur, resulting in the observed insulating state. In the Mott insulator scenario, the localized electron in the 2$a_1$ orbital carries an unpaired spin, resulting in a local magnetic moment of $S = 1/2$. This expectation is supported by the magnetic susceptibility data, showing paramagnetic behavior in the $\alpha$ phase [Fig. 2(f)]. The paramagnetic susceptibility is well fitted to the Curie-Weiss function, yielding an effective paramagnetic Bohr magneton of $p_{\text{eff}} = 1.65$ $\mu_B$, which is consistent with the theoretical value of 1.73 $\mu_B$ for $S = 1/2$.

To account for the insulating state in $\alpha$-Nb$_3$Cl$_8$, electron correlation effects have been considered in previous spin-polarized DFT calculations, where long-range ferromagnetic or antiferromagnetic order was assumed [29,40,41]. However, the resulting magnetically ordered insulating states are inconsistent with the observed paramagnetic behavior in $\alpha$-Nb$_3$Cl$_8$. The static mean-field approximation in DFT + $U$ calculations is inadequate to describe the Mott physics in this material. Therefore, we employed dynamical mean-field theory (DMFT) calculations for this half-filled band within the Hubbard model. This low-energy physical model is justified because the half-filled 2$a_1$ band is isolated from other bands, which have been well described within the single-particle approximation.

In the single-band Hubbard model, the electronic states undergo significant changes with increasing $U$ [Fig. 3(b)]. When $U$ exceeds 0.6 eV, the coherent part at $E_F$ diminishes to negligible levels, while the incoherent part becomes well separated from $E_F$. This indicates that strong correlations drive a Mott transition, splitting the half-filled metallic band into LHB and UHB, with $E_F$ located within the Mott gap between them. In Mott insulators, $U$ determines the magnitude of the splitting between the LHB and UHB, which corresponds to the energy difference between the centers of the two bands [see Fig. S2(c) in Supplemental Material [36]]. The value of $U$ can be estimated using the simple formula $U = \Delta + (W_1 + W_2)/2$, where $\Delta$ is the gap size between the top of the LHB and the bottom of the UHB, and $W_1$ and $W_2$ are the widths of the LHB and UHB, respectively. The PL spectrum in Fig. 2(d) has identified $\Delta$ as 1.0 eV. The LHB and UHB should have similar widths since both of them arise from the splitting of the half-filled band [13]. The ARPES data show that $W_1$ is about 0.2 eV, and therefore we assume that $W_2$ is also about 0.2 eV. The estimated value of $U \sim 1.2$ eV based on our experimental results agrees well with the calculated value of $U \sim 1.15$ eV, which takes the non-local Coulomb interactions into account [42].

Figure 3(c) presents the spectral function of the single-band Hubbard model at $U = 1.2$ eV, showing a splitting of the half-filled band into LHB and UHB, with the magnitude of the splitting determined by $U$. The non-interacting half-filled band needs to be renormalized by a factor of 1.7 when compared with the LHB and UHB, which is consistent with the experimental observation. The PL spectrum indicates a Mott gap size of 1.0 eV, while the ARPES data show that the top of the LHB lies 0.7 eV below $E_F$, suggesting that $E_F$ does not coincide with the middle of the Mott gap in actual samples. Similar behavior has been observed in the parent compounds of cuprate superconductors [43-45]. The ARPES results in Fig. 3(d) are nearly perfectly reproduced by the combined band structure from the HSE06 and DMFT calculations in Fig. 3(e). The half-filled single band splits into LHB and UHB due to strong correlations, while other bands are well described within the single-particle approximation, as illustrated in Fig. 3(f). The excellent agreement demonstrates that $\alpha$-Nb$_3$Cl$_8$ is an ideal Mott insulator described by the single-band Hubbard model. In addition, in the Hubbard model, the exchange coupling $J = -4t^2/U$ is estimated to be -2.4 meV, with the nearest-neighbor hopping parameter $t$ estimated to be 0.027 eV within the tight-binding approximation. The $J$ value is comparable to the $k_B\theta$ value of -1.5 meV determined from the magnetic susceptibility data in Fig. 2(f), with $\theta = -17.8$ K representing the Curie-Weiss temperature.

# IV. MOTT INSULATOR STATE IN $\beta$ PHASE

As the temperature decreases, $Nb_3Cl_8$ undergoes a transition from the paramagnetic $\alpha$ phase to the nonmagnetic $\beta$ phase at $T^* \sim 100$ K, accompanied by significant structural changes [29,30]. The existence of a nonmagnetic ground state is unusual, as Mott insulators usually exhibit magnetically ordered states at low temperatures. The nonmagnetic state has been proposed to arise from either spin singlet [29] or $[Nb_3]^{7+}$–$[Nb_3]^{9+}$ charge disproportionation [30]. An essential question is whether the Mott physics remains significant in the nonmagnetic state. To address this, it is necessary to investigate the electronic structure of the $\beta$ phase. However, because of the highly insulating nature at low temperatures, it is not possible to obtain the electronic structure of $\beta$-$Nb_3Cl_8$ with ARPES. Fortunately, the $\alpha$ to $\beta$ phase transition temperature $T^*$ can be tuned from about 100 K to 400 K in the $Nb_3Cl_{8-x}Br_x$ series with increasing $x$ [46]. Therefore, we synthesize $Nb_3Cl_2Br_6$ samples, where $T^*$ is near room temperature [Fig. 4(a)]. This method allows us to investigate the changes in the electronic structure across $T^*$ through temperature-dependent ARPES measurements.

The ARPES data in Fig. 4(d) reveal that the band structure of $\alpha$-$Nb_3Cl_2Br_6$ closely resembles that of $\alpha$-$Nb_3Cl_8$ (see Fig. S7 in Supplemental Material for more details [36]), confirming $\alpha$-$Nb_3Cl_2Br_6$ as a single-band Mott insulator. The most notable change in the band structure is the splitting of the LHB in the $\beta$ phase, as shown in Fig. 4(e). The temperature-dependent data in Fig. 4(b) show that the splitting occurs at $T^*$, confirming that it arises from the phase transition. However, no discernible changes are observed in the core level peak of the Nb 4$s$ orbital across $T^*$ (see Fig. S8 in Supplemental Material [36]), indicating a lack of variation in the valence state of the Nb ions. This observation contradicts the suggested $[Nb_3]^{7+}$–$[Nb_3]^{9+}$ charge disproportionation [30].

Previous studies have reported significant changes in the stacking sequence of $Nb_3Cl_8$ layers between $\alpha$ and $\beta$ phases [29,30,46]. In the $\alpha$ phase, the $Nb_3$ trimers are staggered between adjacent monolayers [Fig. 4(f)], while in the $\beta$ phase, they stack directly on top of each other within each bilayer [Fig. 4(g)]. Previous DFT calculations have suggested that the structural changes effectively enhance the interlayer coupling of the $Nb_3$ trimers [29,47]. To investigate the effects on the Mott insulator state, we perform DMFT calculations for the bilayer of $\beta$-$Nb_3Cl_8$, considering the interlayer coupling $t_\perp$, while assuming that $U$ remains constant across the phase transition. The value of $t_\perp$ is estimated to be 0.1 eV, which is half of the splitting at the $\Gamma$ point calculated by DFT [see Fig. S10(b) in Supplemental Material [36]]. We note that substituting Cl with Br does not significantly affect the interlayer coupling strength [47]. As shown in Fig. 4(c), the DMFT calculations reveal an energy splitting of both LHB and UHB in the bilayer of $\beta$-$Nb_3Cl_8$, in good agreement with the ARPES data in Fig. 4(e). Therefore, the splitting arises from the enhanced interlayer hybridization, leading to the formation of bonding-antibonding states in the $\beta$ phase.

The strong interlayer hybridization within the bilayer in the nonmagnetic $\beta$ phase is reminiscent of the observation in $VO_2$, where a nonmagnetic insulating phase emerges when the V atoms dimerize [4,5,48]. However, it remains unclear whether the insulating gap in $VO_2$ arises from bonding-antibonding splitting [16,17] or electron correlations [5,18-20], i.e., a Peierls-type band insulator or a Mott insulator. Our ARPES data reveal in $\beta$-$Nb_3Cl_2Br_6$ that the splitting magnitude of 0.26 eV in the LHB is much smaller than the Mott gap size of 1.0 eV. It is evident that the bonding-antibonding splitting has minor effects on the Mott gap, and consequently, the insulating gap is dominantly determined by $U$, as illustrated in Fig. 4(i). Therefore, the $\beta$ phase still falls within the Mott insulator regime.

The nonmagnetic behavior suggests that the spin-1/2 magnetic moments are quenched in the $\beta$ phase. In the on-top stacked $Nb_3$ trimers, a pair of spins is likely to couple antiferromagnetically, forming an interlayer spin singlet in the bonding-antibonding states, as illustrated in Fig. 4(i). This results in a spin-singlet Mott insulator state that exhibits nonmagnetic behavior. Therefore, despite the drastic changes in the magnetic properties, the Mott physics plays a dominant role in both the $\alpha$ and $\beta$ phases.

## V. SUMMARY AND OUTLOOK

Our results firmly establish $\alpha$-$Nb_3Cl_8$ as a canonical spin-1/2 Mott insulator arising from a half-filled flat band with $U \gg W$. Since this band, derived from a single $2a_1$ orbital, is isolated from the high-energy bands, the Mott insulator state can be ideally described by the single-band Hubbard model. The localization of electrons in the $Nb_3$ trimers, due to the long distance between them, suppresses the electronic kinetic energy, leading to a Mott transition. This transition straightforwardly manifests the Mott physics envisioned in Mott's thought experiment [2].

Given the numerous intriguing physical phenomena that have emerged from Mott insulators, there is a strong motivation to manipulate the Mott insulator state in $Nb_3Cl_8$, as well as in related systems like $Nb_3Br_8$ and $Nb_3I_8$, through various methods such as doping, pressure, magnetic fields, and optical pulses. Because of the weak vdW coupling, $Nb_3Cl_8$ crystals can be easily exfoliated into few-layer or even monolayer flakes [35,39], making it convenient to study and utilize the Mott insulator state [49,50]. For instance, a magnetic-field-free Josephson diode has been recently achieved in a vdW heterostructure based on its sister material $Nb_3Br_8$ [51,52]. It is feasible and desirable to fabricate Mott insulator-based devices with this family of materials to produce more exotic quantum phenomena. A previous study has revealed that the transition from the $\alpha$ to $\beta$ phase is entirely suppressed in thin flakes, resulting in the paramagnetic state persisting down to 2 K [30]. While the calculations suggest a 120º antiferromagnetic ground state in the monolayer [42], geometric frustration in the triangular lattices formed by the spin-1/2 $Nb_3$ trimers may lead to more exotic ground states, such as a quantum spin liquid or resonating valence bond state. Lastly, flat-band systems, characterized by a high density of states at $E_F$ and quenched electronic kinetic energy, are considered a fertile ground for strongly correlated phases of matter [41,53-55]. The Mott insulator state derived from a single flat band makes $Nb_3Cl_8$ a paradigmatic system that exemplifies strongly correlated physical phenomena resulting from flat-band effects. Our discovery will stimulate intensive exploration of the flat-band effects in the materials database.

## Acknowledgements

We thank X.-Z. Chen, B. Jiang, M.-Z. Hu, F.-M. Chen, J.-D. Liu, I. Vobornik, J. Fujii, N. C. Plumb, A. Pfister, H.-E. Zhu, C. Polley, and N. Olszowska for technical assistance. This work was supported by the Ministry of Science and Technology of China (Grants No. 2022YFA1403800, No. 2018YFA0305700, No. 2018YFE0202600, No. 2020YFA0308800, and No. 2022YFA1403401), the National Natural Science Foundation of China (Grants No. U22A6005, No. U1832202, No. 11925408, No. 11921004, No. 12188101, No. U2032204, No. 12274459, No. 22073111, No. 92250307, No. 12174428, No. 12061131002, No. 11888101, No. U21A20432, No. 12321004, No. 12025407, and No. 92065109), the Chinese Academy of Sciences (Grants No. QYZDB-SSW-SLH043, No. XDB33000000, and No. XDB28000000), the Informatization Plan of Chinese Academy of Sciences (Grant No. CAS-WX2021SF-0102), the Beijing Natural Science Foundation (Grant No. Z210006 and No. Z200005), the China Scholarship Council (Grant No. 202104910090), the Sino-Swiss Science and Technology Cooperation (Grant No. CN-EG-03-012021), the Swiss National Science Foundation (Grant No. 200021-188413), the Fundamental Research Funds for the Central Universities and Research Funds of Renmin University of China (RUC) (Grants No. 18XNLG14, No. 19XNLG13, and No. 19XNLG17), the Outstanding Innovative Talents Cultivation Funded Programs 2022 of Renmin University of China, and the Beijing National Laboratory for Condensed Matter Physics. We acknowledge the Synergetic Extreme Condition User Facility (SECUF), the "Dreamline" beamline of Shanghai Synchrotron Radiation Facility (SSRF), the Steady High Magnetic Field Facility (SHMFF), the SIS-ULTRA beamline of the Swiss Light Source, the APE-LE beamline of the Elettra Light Source (Proposal No. 20215821), the 13U ARPES beamline of the National Synchrotron Radiation Laboratory in Hefei (Proposal No. 2021-HLS-PT-004344), the "Bloch" beamline of MAXIV in Sweden (Proposal No. 20200353), and the "UARPES" beamline of SOLARIS in Poland (Proposal No. 211070).

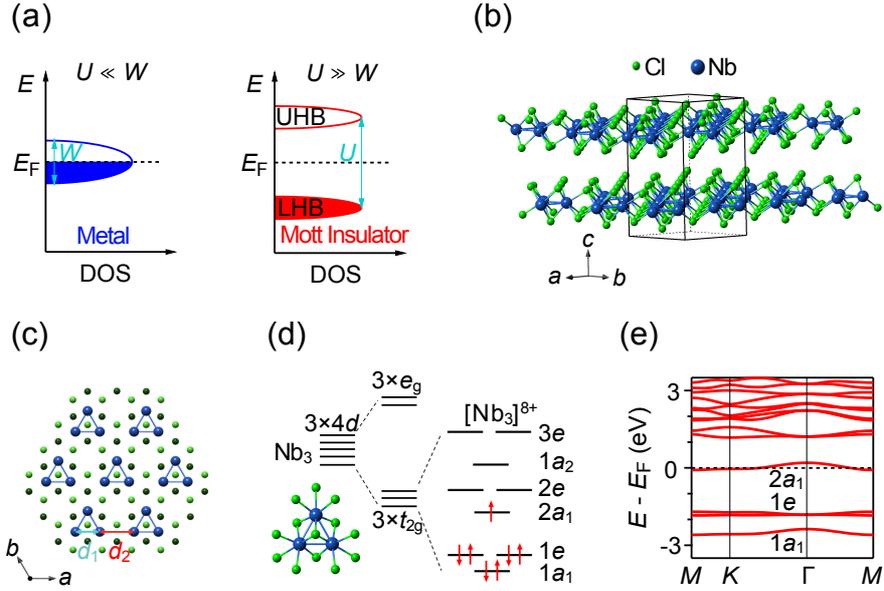

Fig. 1. Half-filled flat band in the monolayer of $\alpha$-Nb$_3$Cl$_8$. (a) Schematic illustrations of a metal for $U \ll W$ (left) and a Mott insulator for $U \gg W$ (right). (b) Crystal structure of $\alpha$-Nb$_3$Cl$_8$. (c) Top view of the monolayer of Nb$_3$Cl$_8$ showing the Nb$_3$ trimers. Blue lines indicate the Nb–Nb metal bonds within each Nb$_3$ trimer. Dark- and light-green spheres represent the Cl atoms below and above the Nb layer, respectively. (d) Schematic illustration of the molecular orbitals in an isolated Nb$_3$Cl$_{13}$ cluster, occupied by seven Nb 4$d$ valence electrons for a [Nb$_3$]$^{8+}$ trimer. Inset: top view of a Nb$_3$Cl$_{13}$ cluster composed of three edge-sharing NbCl$_6$ octahedra. (e) Calculated band structure of the monolayer of Nb$_3$Cl$_8$ using the HSE06 functional.

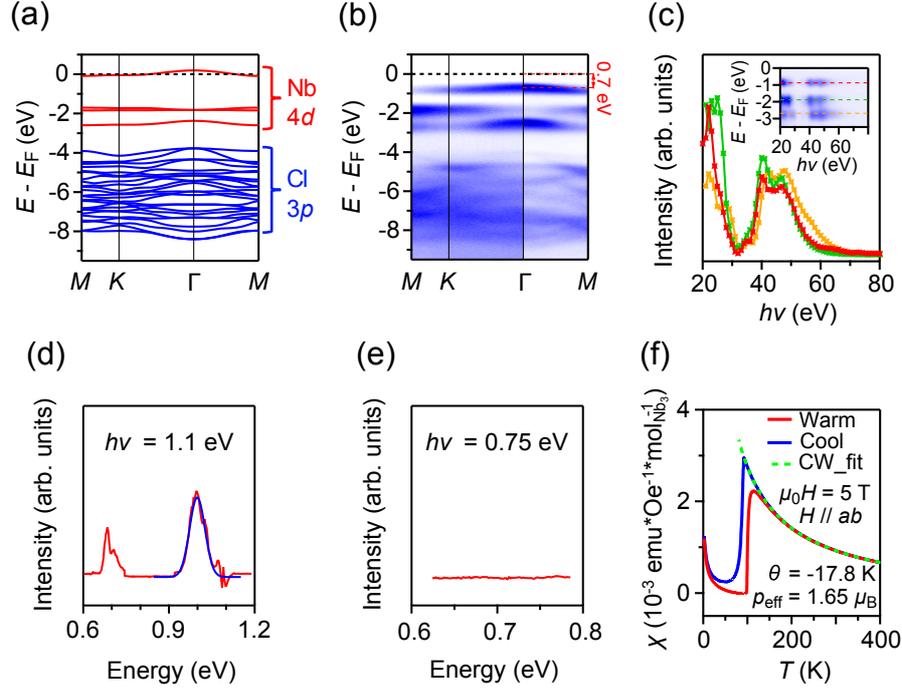

Fig. 2. Band gap at $E_F$ revealed by ARPES and PL spectra. (a) Calculated valence band structure of the monolayer of $Nb_3Cl_8$ using the HSE06 functional. Red and blue curves represent the bands originating from the Nb 4$d$ and Cl 3$p$ orbitals, respectively. (b) Intensity plot of the ARPES data along $M$–$K$–$\Gamma$–$M$. (c) $h\nu$ dependence of the spectral intensities at three constant energies indicated by the dashed lines in the inset. Inset: intensity plot of the ARPES data at the $M$ point with varying $h\nu$. The spectral intensities are normalized by photon flux. (d) PL spectrum excited by 1.1-eV photons (red curve). Stray signals caused by the incident light have been subtracted (see Fig. S3 in Supplemental Material [36]). The peak at 1.0 eV is fitted to a Gaussian function (blue curve). (e) Same as in (d) but excited by 0.75-eV photons. (f) Temperature-dependent magnetic susceptibility measured upon cooling (blue) and warming (red). The green dashed curve represents a fitting of the cooling data from 100 to 400 K to the Curie-Weiss function.

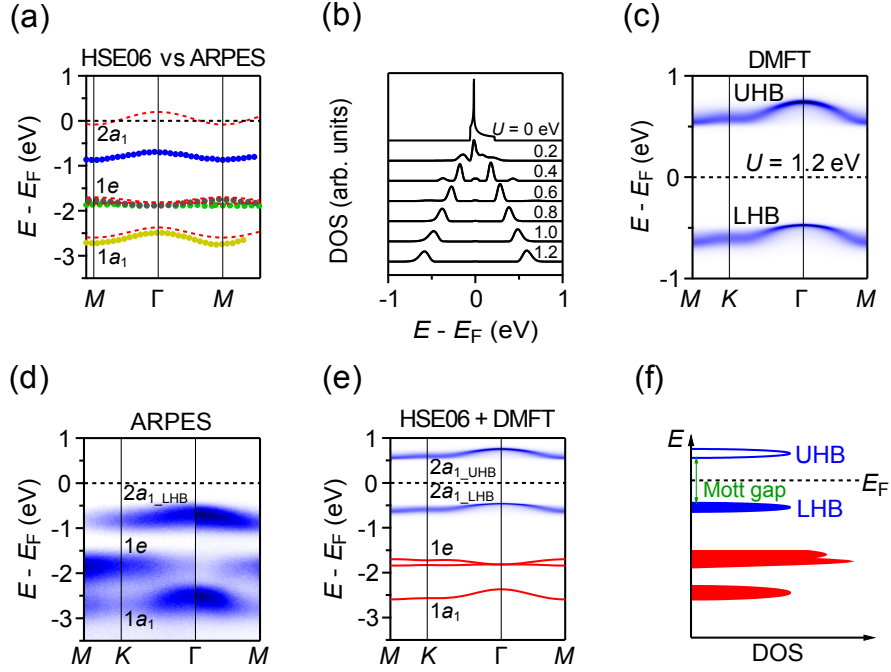

Fig. 3. Single-band Mott insulator state in $\alpha$-$Nb_3Cl_8$. (a) Comparison between the calculated and experimental bands. Red dashed curves represent the calculated bands based on the HSE06 functional. The experimental bands are extracted by tracking the peak positions of energy distribution curves in the ARPES data. Raw ARPES data are shown in Figs. S5(a) and S5(b) in the Supplemental Material [36]. (b) Density of states (DOS) of the single-band Hubbard model with increasing $U$ from 0 to 1.2 eV. (c) Spectral function of the LHB and UHB along the high-symmetry path with $U = 1.2$ eV. (d) Intensity plot of the ARPES data along $M–K–\Gamma–M$. (e) Combined band structure from the HSE06 (red) and DMFT (blue) calculations, which nearly perfectly reproduces the ARPES data in (d). (f) Schematic illustration of the electronic structure of $\alpha$-$Nb_3Cl_8$, showing a single-band Mott insulator state.

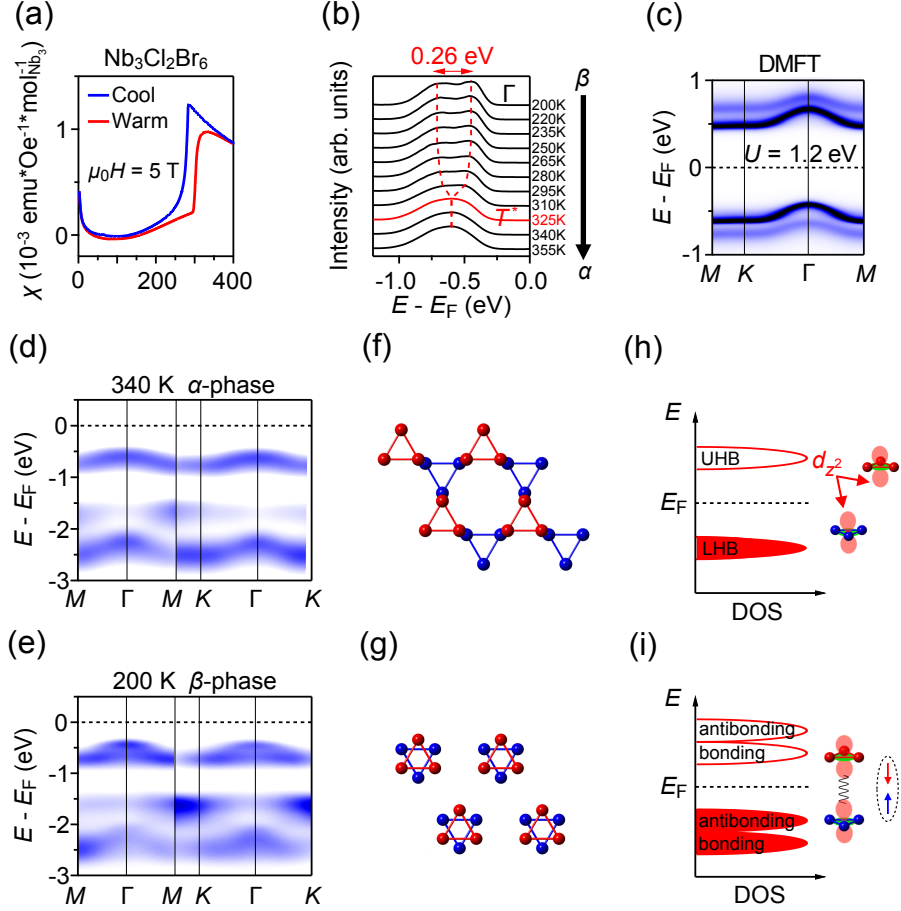

Fig. 4. Electronic structure changes between the $\alpha$ and $\beta$ phases in Nb$_3$Cl$_2$Br$_6$. (a) Temperature-dependent magnetic susceptibility measured upon cooling (blue) and warming (red). (b) Energy distribution curves at the $\Gamma$ point measured at different temperatures upon warming. Red dashed curves are guides to the eye for the peak positions, showing a splitting of the LHB below $T^*$. (c) Spectral function of the bilayer of $\beta$-Nb$_3$Cl$_8$ calculated by DMFT with $U = 1.2$ eV and $t_\perp = 0.1$ eV, showing the splitting of both LHB and UHB. (d,e) Intensity plots of the ARPES data along $\Gamma$–$M$–$K$–$\Gamma$ collected at 340 and 200 K, respectively. (f,g) Top view of the Nb$_3$ trimers of one bilayer in the $\alpha$ and $\beta$ phases, respectively. Red and blue circles indicate the Nb ions in the top and bottom monolayers, respectively. (h,i) Schematic illustrations of the electronic structures in the $\alpha$ and $\beta$ phases, respectively. The insets show the $d_{z^2}$ orbitals of the Nb$_3$ trimers in the bilayer, which are staggered in the $\alpha$ phase (h), but directly on-top in the $\beta$ phase to form an interlayer spin singlet (i).

# Supplemental Material for

# Discovery of a single-band Mott insulator in a van der Waals flat-band compound

Shunye Gao,[#] Shuai Zhang,[#] Cuixiang Wang,[#] Shaohua Yan,[#] Xin Han, Xuecong Ji, Wei Tao, Jingtong Liu, Tiantian Wang, Shuaikang Yuan, Gexing Qu, Ziyan Chen, Yongzhao Zhang, Jierui Huang, Mojun Pan, Shiyu Peng, Yong Hu, Hang Li, Yaobo Huang, Hui Zhou, Sheng Meng, Liu Yang, Zhiwei Wang, Yugui Yao, Zhiguo Chen, Ming Shi, Hong Ding, Huaixin Yang, Kun Jiang, Yunliang Li, Hechang Lei,[*] Youguo Shi,[*] Hongming Weng,[*] Tian Qian,[*]

[#] These authors contributed equally to this work.

[*] Corresponding authors: tqian@iphy.ac.cn, hmweng@iphy.ac.cn, ygshi@iphy.ac.cn, hlei@ruc.edu.cn

This file includes:

   Materials and Methods

   Figures S1 to S10

   References [1-15]

# Materials and Methods

**Part I. Sample crystal growth**

Single crystals of $Nb_3Cl_8$ were grown using $PbCl_2$ as the flux. The starting materials, Nb (Alfa Aesar, 99.99 %) and $NbCl_5$ (Alfa Aesar, 99.9 %), were mixed with a molar ratio of 7:8 in a glovebox filled by argon. The mixture was then placed in an alumina crucible and sealed in quartz tube under vacuum. The entire setup was heated to 750 °C for 150 hours and then cooled to room temperature. Finally, the single crystals were separated from the $PbCl_2$ flux using hot water.

Single crystals of $Nb_3Cl_2Br_6$ were grown using the chemical vapor transport method. A mixture of Nb (99.95%), $NbCl_5$ (99.95%), and $NbBr_5$ (99.9%) in a 7:2:6 molar ratio was placed into a silicon tube with 20 mg $NH_4Br$ (99.99%). The tube was evacuated down to 0.01 Pa and sealed under vacuum, then positioned in a two-zone horizontal tube furnace. The two growth zones were slowly heated to 840 °C and 785 °C for 48 hours and then held at these temperatures for another 120 hours. Subsequently, the furnace was cooled naturally. Shiny, plate-like crystals with lateral dimensions of up to several millimeters were obtained from the growth.

**Part II. Angle-resolved photoemission spectroscopy experiments**

The angle-resolved photoemission spectroscopy (ARPES) data were acquired at the "Dreamline" beamline of the Shanghai Synchrotron Radiation Facility and the SIS-ULTRA beamline of the Swiss Light Source. Complementary experiments were conducted at the APE-LE beamline of the Elettra Light Source under proposal 20215821, the 13U ARPES beamline of the National Synchrotron Radiation Laboratory at Hefei under proposal 2021-HLS-PT-004344, the "Bloch" beamline of MAXIV in Sweden under proposal 20200353, and the "UARPES" beamline of SOLARIS in Poland under proposal 211070. The samples were cleaved *in situ* and measured at a photon energy of $h\nu = 40$ eV except for $h\nu$ dependent measurements. The Fermi level ($E_F$) was calibrated using the energy position of the Fermi edge of gold.

**Part III. Photoluminescence spectroscopy experiments**

1 KHz pulse train with a duration of 90 fs centered at 800 nm and an output of 3 mJ was generated using a Ti:sapphire regenerative amplifier (Spectra-Physic, Spitfire) seeded with an oscillator (Spectra-Physics, Spitfire). Approximately 70% of the output was directed towards pumping a commercially automated optical parametric amplifier (TOPAS, Spectra-Physic, Spitfire) capable of producing the infrared (IR) light tunable from 1.1 to 2.6 um with a pulse energy of 200 uJ. The wavelength range of 800 nm to 1 um was further generated by transmitting the IR pulse through a beta-barium borate crystal (BBO). Visible light spanning from 266 to 760 nm was produced with a home-build non-collinear optical parametric amplifier (NOPA) system. The light with the desired wavelength and optimized energy was then focused onto the sample with a spot size of 150 um. The scattered signal from the sample surface was collected by a 64-channel Mid-IR detector (FPAS 0144 Infrared Systems Development) in combination with a 300 mm focal length spectrometer (Princeton Instrument 2300i). To reveal the band gap, the samples were symmetrically irradiated with different central wavelength varying from 400 nm to 1.5 um.

**Part IV. First-principles calculations**

Electronic structure calculations were performed using the Vienna *Ab initio* Simulation Package (VASP) [1]. The projector-augmented-wave (PAW) [2,3] method with the Heyd-Scuseria-Ernzerhof (HSE06) hybrid functional [4,5] was employed. A plane-wave cutoff of 500 eV was set for the kinetic energy, and a $6 \times 6 \times 1$ Γ-centered Monkhorst-Pack *k*-point mesh was used to sample the Brillouin zone during the self-consistent charge convergence calculations. For a more efficiently calculation of the density of states (DOS), Wannier functions were constructed for bands composed of the Nb 4*d* and Cl 3*p* orbitals using the Wannier90 package [6]. The crystal structure used in the calculations was obtained from Ref. [7].

**Part V. Dynamical mean-field theory calculations**

We performed the dynamical mean-field theory (DMFT) calculations [8] on the single-band Hubbard model using the IQIST package [9], which utilizes the continuous-time quantum Monte Carlo (CTQMC) [10] impurity solver to determine the impurity imaginary time Green's function. Due to the absence of magnetic order in this material, we focused on the nonmagnetic phase in the DMFT calculations and neglected the spin-orbital coupling effect for simplicity. The DOS obtained from the single band crossing $E_F$ was used to initialize the local Green's function. The DMFT calculations were performed at a temperature controlled by the parameter $\beta$ of 400 corresponding to approximately 29.01 K. The spectral function was obtained from the analytic continuation using the maximal entropy method and Pade approximation [9,11].

**Part VI. Transmission electron microscopy experiments**

We conducted transmission electron microscopy measurements on a multilayer $Nb_3Cl_2Br_6$ film using JEM2100F to obtain convergent beam electron diffractions at 300 and 96 K, respectively. The film was fabricated by mechanically exfoliating a bulk single crystal with the [001] direction parallel to the surface normal.

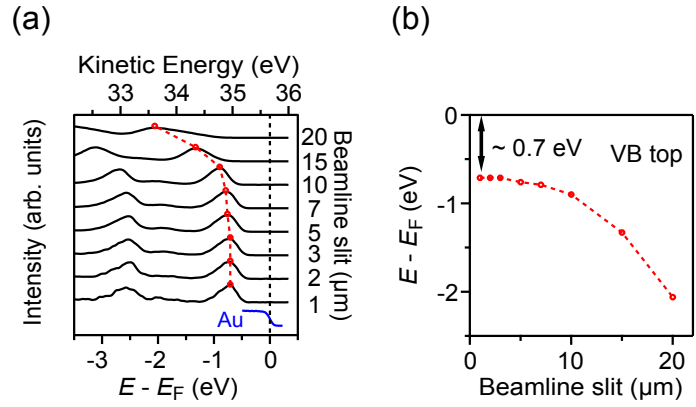

Fig. S1. Charging effects in the ARPES spectra of a $Nb_3Cl_8$ sample at room temperature. (a) Energy distribution curves (EDCs) at the $\Gamma$ point of $Nb_3Cl_8$ obtained with different beamline slit sizes, which are proportional to photon flux. Blue curve is the EDC of gold. All the data were taken at $h\nu = 40$ eV. (b) Energy position of the peak in the EDCs, corresponding to the valence band (VB) top at the $\Gamma$ point, plotted as a function of beamline slit size.

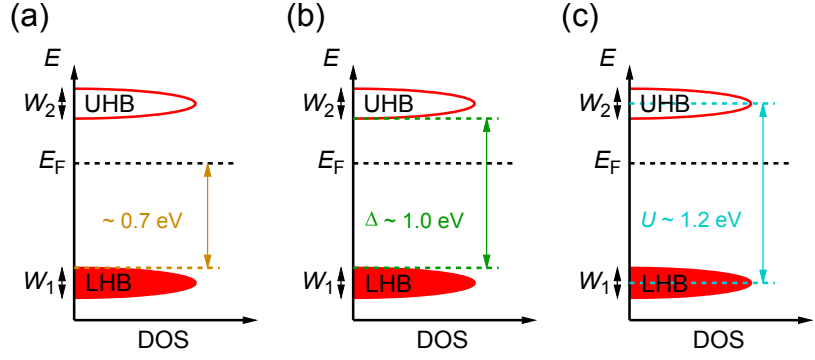

Fig. S2. Schematic illustrations of (a) the gap between the top of the lower Hubbard band (LHB) and $E_F$ observed in APRES experiments, (b) the gap $\Delta$ between the top of the LHB and the bottom of the upper Hubbard band (UHB) determined from photoluminescence (PL) experiments, and (c) the correlation strength $U$ corresponding to the energy difference between the centers of the LHB and UHB in $\alpha$-Nb$_3$Cl$_8$. $W_1$ and $W_2$ represent the widths of the LHB and UHB, respectively.

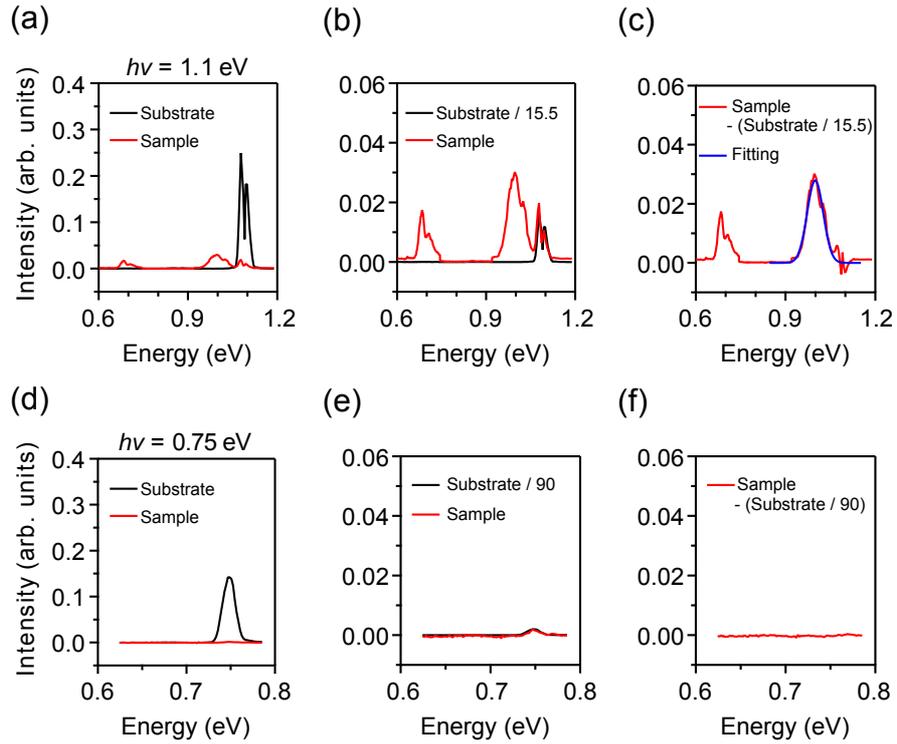

Fig. S3. (a) PL spectrum of the Nb$_3$Cl$_8$ sample (red curve) excited by 1.1 eV photons. To remove the stray signals from the incident light in the PL spectrum, we also measured the spectrum of the substrate (black curve). (b) The spectral intensity of the substrate is normalized to the stray signals in the spectrum of the sample. (c) PL spectrum of the sample with the stray signals subtracted. Blue curve is a fitting of the peak at 1.0 eV using a Gaussian function. (d)-(f) Same as (a)-(c) but excited by 0.75 eV photons. All the data were collected at room temperature.

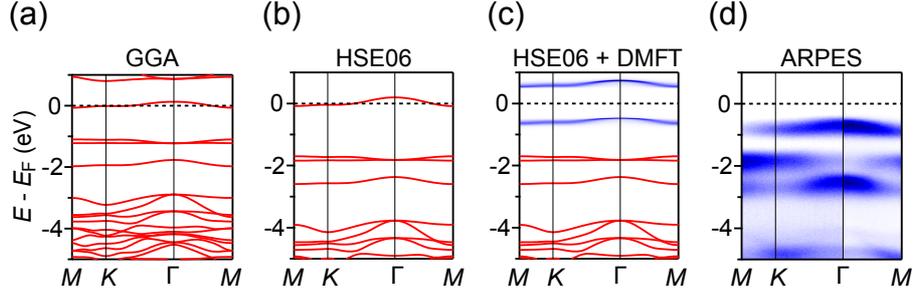

Fig. S4. Comparison between the calculated band structures for the monolayer of Nb$_3$Cl$_8$ and the ARPES results of $\alpha$-Nb$_3$Cl$_8$. (a)-(c) Calculated band structures using generalized gradient approximation (GGA), HSE06, and HSE06 + DMFT, respectively. (d) Intensity plot of the ARPES data along $\Gamma-M-K-\Gamma$ of $\alpha$-Nb$_3$Cl$_8$. Generally, the local density approximation or GGA tends to underestimate band gaps due to the derivative discontinuity of the exchange-correlation potential [12,13]. To address this issue, the band structure calculated using the HSE06 hybrid functional is employed in the discussion [14,15].

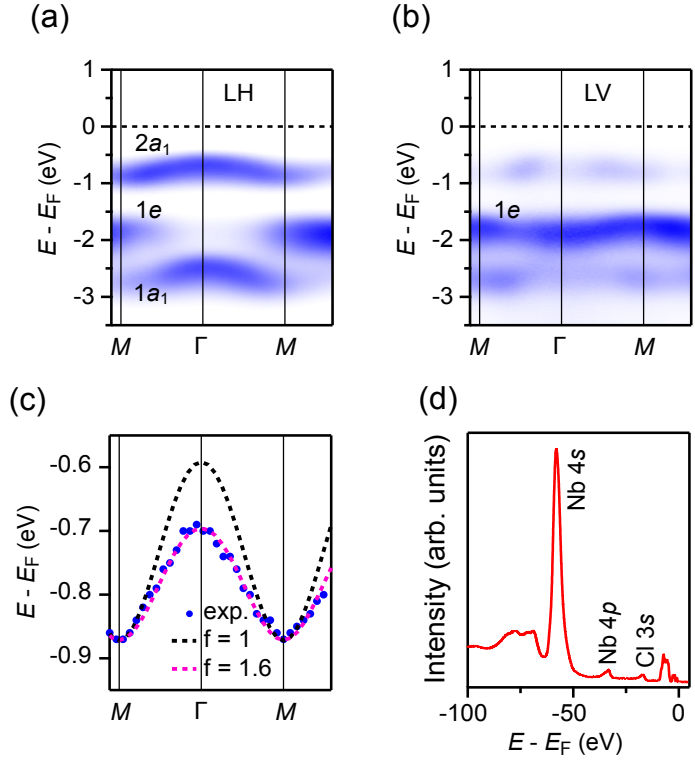

Fig. S5. (a),(b) Intensity plots of the ARPES data of $\alpha$-Nb$_3$Cl$_8$ along $\Gamma-M$ collected with linearly horizontally (LH) and vertically (LV) polarized light, respectively. Two almost degenerate 1$e$ bands are observed with LH and LV polarized light, respectively. (c) Precise comparison between the experimental and calculated 2$a_1$ bands. The calculated band is shift down by 0.8 eV. Black and pink curves represent the calculated 2$a_1$ bands, which are unrenormalized and renormalized by a factor of 1.6, respectively. (d) Core-level photoemission spectrum of $\alpha$-Nb$_3$Cl$_8$ measured at $h\nu = 240$ eV, showing characteristic peaks of the Nb 4$s$, Nb 4$p$, and Cl 3$s$ orbitals.

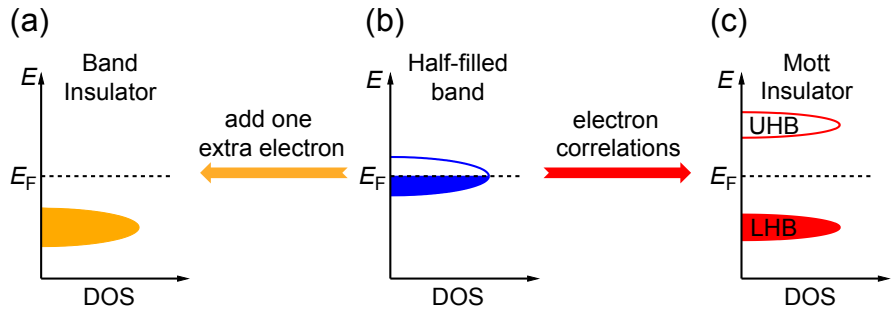

Fig. S6. Schematic plots of (a) a band insulator and (c) a Mott insulator derived from (b) a metal with a half-filled band. The band insulator is obtained by adding one extra electron to fill this band, whereas the Mott insulator emerges due to strong electron correlations.

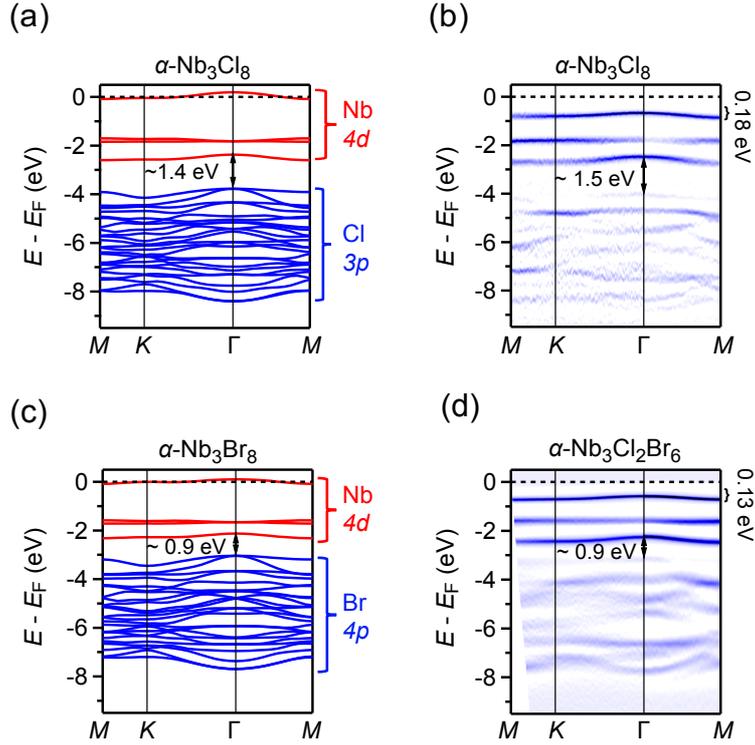

Fig. S7. (a),(b) Calculated band structures using the HSE06 functional for the monolayers of $Nb_3Cl_8$ and $Nb_3Br_8$, respectively. (c),(d) Intensity plots of the curvature with respect to energy of the ARPES data of $\alpha$-$Nb_3Cl_8$ and $\alpha$-$Nb_3Cl_2Br_6$, respectively. The gap size of two sets of bands, originating from the Nd 4$d$ orbitals and the ligand $p$ orbitals, respectively, decreases when substituting Cl with Br, which is attributed to the lower electronegativity of Br compared to Cl. In addition, the substitution increases the inter-cluster Nb–Nb distance, which further suppresses the electronic kinetic energy, resulting in a narrower width of the half-filled band in $Nb_3Cl_2Br_6$ compared to $Nb_3Cl_8$.

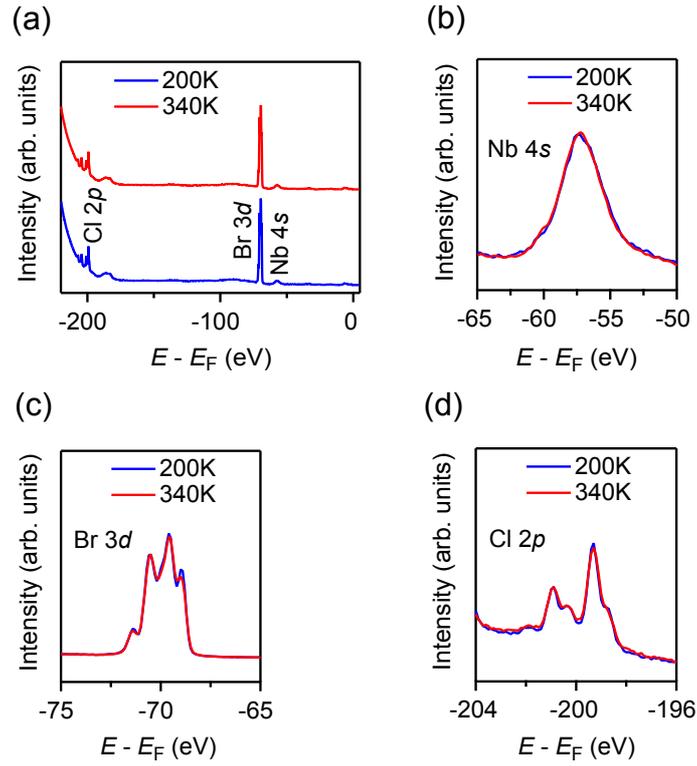

Fig. S8. (a) Core-level photoemission spectra of $Nb_3Cl_2Br_6$ at 340 K ($\alpha$ phase) and 200 K ($\beta$ phase). (b)-(d) Core-level photoemission spectra in a narrow energy range, showing characteristic peaks of the Nb 4*s*, Br 3*d*, and Cl 2*p* orbitals, respectively.

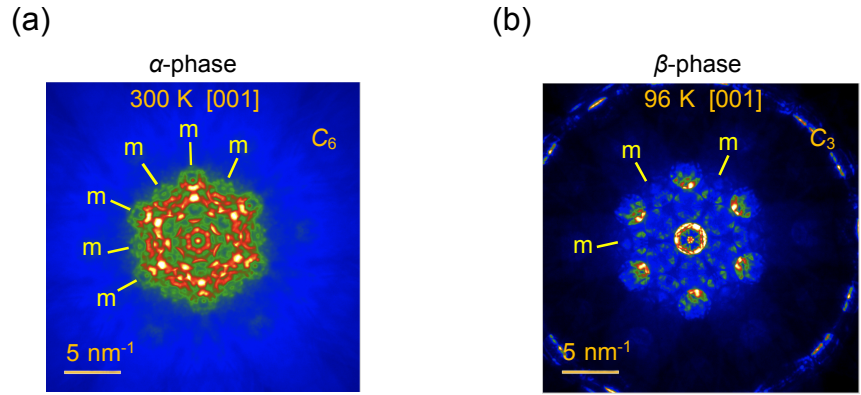

Fig. S9. Convergent beam electron diffraction patterns measured by transmission electron microscopy under the zone axis [001] at 300 K ($\alpha$ phase) and 96 K ($\beta$ phase), respectively, for a $Nb_3Cl_2Br_6$ sample. Yellow lines indicate the reflection planes. The results reveal that as the temperature decreases from 300 K to 96 K (transition from $\alpha$ to $\beta$ phase), the symmetry element along the [001] direction changes from $C_6$ to $C_3$, which is consistent with the previously reported structure adopting the $R3$ space-group symmetry in the $\beta$ phase [7].

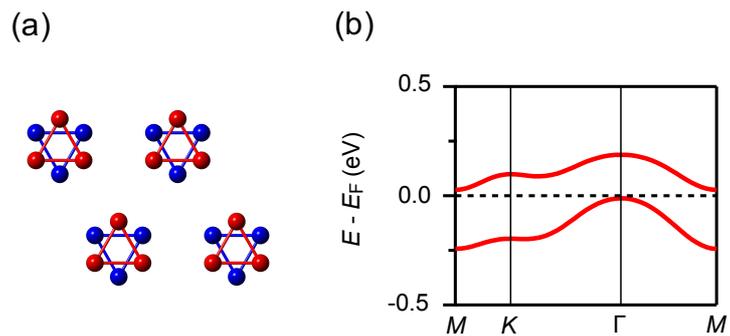

Fig. S10. (a) Crystal structure of one bilayer of $\beta$-Nb$_3$Cl$_8$. For simplicity, only Nb ions are shown. Red and blue circles indicate the Nb ions in the top and bottom monolayers, respectively. (b) DFT calculated band structure near $E_F$ for the bilayer of $\beta$-Nb$_3$Cl$_8$.